\documentclass[twocolumn]{article}    
\pdfoutput=1

\usepackage{graphicx}          
\usepackage{amsmath}
\usepackage{array}
\usepackage{comment}
\usepackage{graphicx}
\usepackage{caption}
\usepackage{subfigure}
\usepackage{cases}
\usepackage{amssymb}
\usepackage{enumerate}
\usepackage{cite}
\usepackage[numbers,sort&compress]{natbib}
\usepackage{wrapfig}
\usepackage{graphicx}
\newtheorem{lemma}{Lemma}
\newtheorem{theorem}{Theorem}

\newtheorem{remark}{Remark}
\newtheorem{definition}{Definition}

\usepackage[colorlinks,
linkcolor=blue,
anchorcolor=blue,
citecolor=blue
]{hyperref}
\usepackage{color}
\makeatletter
\renewcommand{\maketag@@@}[1]{\hbox{\m@th\normalsize\normalfont#1}}%
\makeatother
\usepackage{authblk}

\title{The stability of bi-polarization on dynamical directed graphs: an emergent game perspective}

\author[a,b]{Yakun Wang}    
\author[a,b,c]{Yuan Liu}               
\author[a,b]{Bin Wu \thanks{\href{mailto:bin.wu@bupt.edu.en}{bin.wu@bupt.edu.en}}}  
\affil[a]{School of Science, Beijing University of Posts and Telecommunications, Beijing 100876, China}  
\affil[b]{Key Laboratory of Mathematics and Information Networks (Beijing University of Posts and Telecommunications), Ministry of Education, Beijing 100876, China}             
\affil[c]{Department of Theoretical Biology, Max Planck Institute for Evolutionary Biology, Plön 24306, Germany} 

\begin{document}


\maketitle 


%
%

\begin{abstract}                          
This paper proposes a co-evolutionary model of directed graphs and three opinions, i.e., conservative$(+)$, neutral$(\odot)$ and liberal$(-)$. Agents update both opinions and social relationships with bias. We find that an emergent game suffices to predict the stability of bi-polarization under a rare opinion updating limit and a large system size limit. The bi-polarization is stable if and only if the emergent game has an internal Nash equilibrium. The necessary and sufficient condition is explained by both risk dominance and evolutionary stability. This game approach facilitates us to reveal the stability of bi-polarization in empirical systems. Our work fosters the understanding of opinion formation for controversial topics, and shows a deep connection between opinion dynamics and evolutionary game theory.
\end{abstract}


\section{Introduction}
\begin{figure}
	\centering
	\includegraphics[scale=0.13]{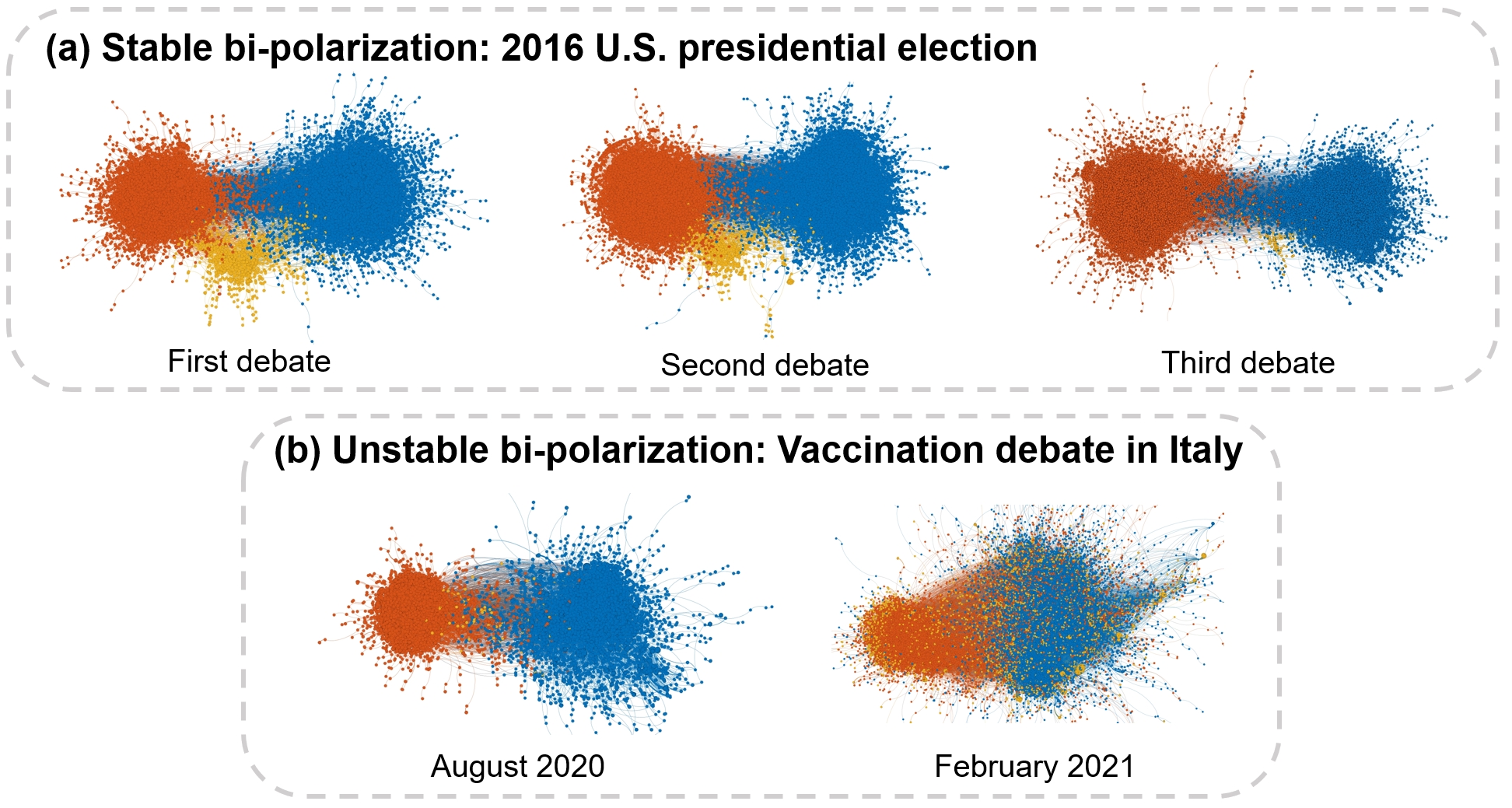}
	\caption{\textbf{Stable bi-polarization and unstable bi-polarization on Twitter.} (a) Stable bi-polarization in the 2016 U.S. presidential election \cite{PRX_echo_chamber}. Red is conservative, blue is liberal and yellow is neutral. The fraction of neutral opinion decreases and the changes of extreme opinions are not significant within the three debates. (b) Unstable bi-polarization in the Italian vaccination debate \cite{Vaccine_Lachi_twitter}. Red is Pro-vaccine, blue is No-vaccine and yellow is neutral. The fraction of neutral supporters vary widely. }
	\label{stable_bipolarization}
\end{figure}

Opinion dynamics has received much more attention over the past decades \cite{ARC_opinion_01, ARC_opinion_02, Anahita_opinion_01}.
Controversial topics lead to bi-polarization, that is, agents with opposite opinions are divided into two groups \cite{Automatica_polarization, Echo_chamber_PNAS_2021, PRL_Modeling}. 
In social media, both stable bi-polarization and unstable bi-polarization are found (see Fig. \hyperref[stable_bipolarization]{1}). 
For example, stable bi-polarization is in the 2016 U.S. presidential election on Twitter (now ``X") \cite{PRX_echo_chamber}. 
Unstable bi-polarization is in the COVID-19 outbreak in the Italian vaccination debates on Twitter \cite{Vaccine_Lachi_twitter}. 
It is not clear why some topics evolve to stable bi-polarization, and why others do not. 
It brings up a crucial question in opinion dynamics: what is the driving force of the stability of bi-polarization? 

Either necessary or sufficient conditions for stable bi-polarization are shown.
If agents update their opinions by reinforcement learning, then bi-polarization emerges \cite{polarization_stable_JMS}. But this bi-polarization is unstable such that agents reach consensus asymptotically almost surely \cite{polarization_stable_02}. 
Therefore reinforcement learning is \emph{not} sufficient for stable bi-polarization.
Homophily without bias assimilation does \emph{not} lead to bi-polarization, which suggests that bias assimilation is necessary \cite{Dandekar_PNAS_2012}. Weakly biased assimilation leads to an unstable bi-polarization state. 
And bi-polarization is stable if agents have the intermediate or strong bias assimilation \cite{Automatica_polarization}.
It implies that intermediate or strong bias assimilation is necessary for stable bi-polarization.

All of the above stability of bi-polarization results are based on static graphs, on which agents have fixed interactions over time.
This is in contrast with the real social systems,
where relationships change over time.
In addition, we note 
i) social interactions are captured by directed graphs,
mirroring the leader-follower dynamics in multi-agent systems;
ii) agents prefer to learn the similar opinions, i.e., bounded confidence.
Motivated by these, we propose a co-evolving model of opinions and directed graphs. 
An emergent game facilitates us to give a necessary and sufficient condition for stable bi-polarization, provided that the directed graph evolves much faster than the opinions. Empirical data validates the necessary and sufficient condition. 
Our work not only gives the driving force of how social graphs should be adjusted for stable bi-polarization, but also sheds a deep connection between stable bi-polarization and evolutionary games.

The paper is organized as follows. Section \hyperref[coevolutionary_model]{2} introduces the co-evolution model. Section \hyperref[an_emergent_game]{3} analyzes the stability of bi-polarization via an emergent game. Section \hyperref[simulations]{4} validates the necessary and sufficient condition by empirical data from an emergent game perspective. Conclusions are found in Section \hyperref[empirical_validation]{5}. 

\section{Co-evolutionary model}
\label{coevolutionary_model}
The social relationships are captured by a directed graph  by $\mathcal G(t)=(\mathcal V,\mathcal E(t))$, where $\mathcal V = \{1,2, \cdots N\}$ is the set of agents and $\mathcal E(t) = \{(i,j): i,j \in \mathcal V\}$ is the set of directed links at time t. $A(t) = {\{ {A_{ij}}(t)\} _{N \times N}}$ is the adjacency matrix of the graph $\mathcal G(t)$. If agent $i$ is connected to agent $j$ at time $t$, then $(i,j) \in \mathcal E(t)$ and ${A_{ij}}(t) = 1$. Otherwise, $\left( {i,j} \right) \notin \mathcal E(t)$ and ${A_{ij}}(t) = 0$. Denote $T_i(t) = \{ j \in V:{A_{ij}}(t) = 1\} $ as the teacher set of agent $i$. Denote $S_i(t) = \{ k \in V:{A_{ki}}(t) = 1\} $ as the student set of agent $i$. Denote $L$ as the average in(out)-degree. There are three opinions, i.e., conservative opinion $(+)$, liberal opinion $(-)$, and neutral opinion $( \odot )$. Denote $P = \left\{ { + , \odot , - } \right\}$ as the opinion set. Denote $x_m$ as the fraction of opinion $m$, where $m \in P$. Each agent $i$ adopts one of the three opinions denoted as $X_i(t)$, where $X_i(t) \in P$. Denote $T_i^m(t) = \{ j \in T_i (t): X_j(t) = m, m \in P\} $ as the set of agents with opinion $m\in P$ in the teacher set of agent $i$, i.e., $T_i(t)$. Denote $[(i,j)]$ as the type of directed link $(i,j)$, where $(i,j) \in \mathcal E(t)$ and $[(i,j)] \in E = \left\{ {\overrightarrow { +  + } ,\overrightarrow { \odot  \odot } ,\overrightarrow { -  - } ,\overrightarrow { +  \odot } ,\overrightarrow { \odot  + } ,\overrightarrow { -  \odot } ,\overrightarrow { \odot  - } ,\overrightarrow { +  - } ,\overrightarrow { -  + } } \right\}$. Denote ${\pi _{[(i,j)]} (t)}$ as the fraction of directed link $[(i,j)]$, where $(i,j) \in \mathcal E(t)$.

\begin{figure}[h]
	\centering
	\includegraphics[scale=0.50]{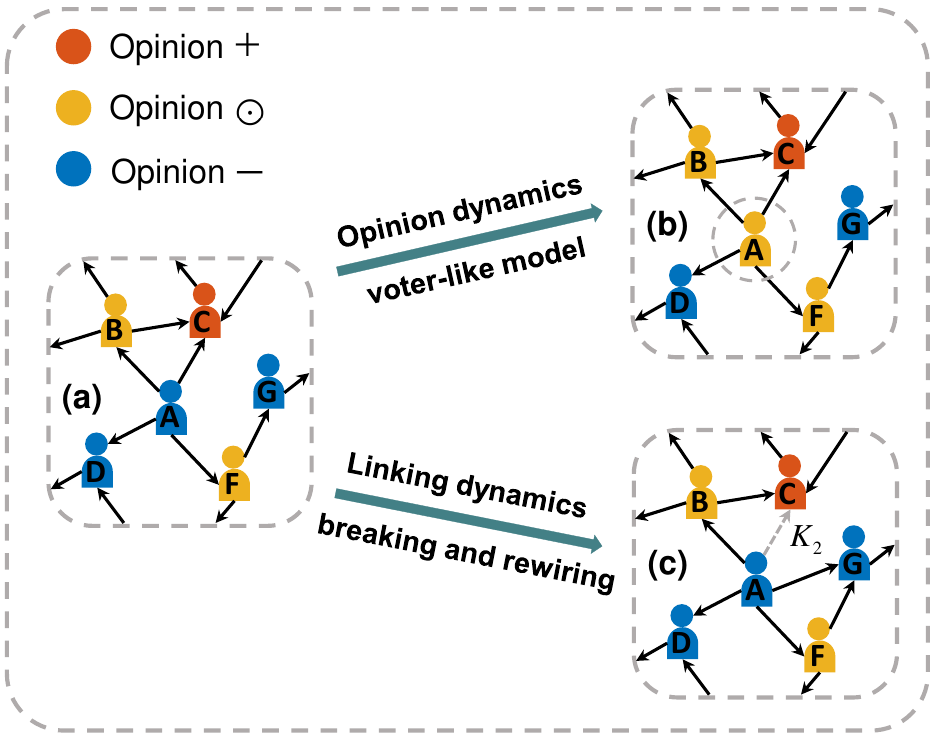}
	\caption{\textbf{Coevolutionary dynamics of opinions and directed social relationships.} \textbf{(a)} The social relationships are captured by directed graphs. \textbf{(b)} With probability $\phi $, opinion updating happens. $A$ with opinion $-$ is selected. And $A$ adopts opinion $\odot$ with probability $1/2$. \textbf{(c)} With probability $1 - \phi $, linking dynamics happens. $\protect\overrightarrow {AC} $ is selected. $A$ is chosen randomly and breaks the directed link $\protect\overrightarrow {AC} $ with probability $K_2$. Then $A$ reconnects to $G$. }
	\label{opinion_and_linking_bi_polarization}
\end{figure} 

\subsection{Opinion dynamics}
Opinion dynamics happens with probability $\phi $ (see Fig. \hyperref[opinion_and_linking_bi_polarization]{2}). An agent $i$ is randomly selected. 
And a teacher $j$ in the teacher set $T_i(t)$ is randomly selected.
If the agent $i$ holds opinion $+$,
and the teacher $j$ holds opinion $-$,
then the agent $i$ does not change her opinion.
If the teacher $j$ holds opinion $+$ or opinion $\odot$, then the agent $i$ adopts teacher's opinion. 
The probability that the agent $i$ holds opinion $\odot$ is the fraction of opinion $\odot$ in the teacher set $T_i(t)$.
The probability that the agent $i$ holds opinion $+$ is the fraction of opinion $+$ and opinion $-$ in the teacher set $T_i(t)$. That is, if $X_i(t) = +$, then 
\begin{equation}
\left\{ \begin{array}{l}
P\left( {{X_i}\left( {t + 1} \right) =  + } \right) = \frac{\phi }{N}\frac{{\left| {T_i^ + \left( t \right)} \right| + \left| {T_i^ - \left( t \right)} \right|}}{{\left| {{T_i}\left( t \right)} \right|}}\\
P\left( {{X_i}\left( {t + 1} \right) =  \odot } \right) = \frac{\phi }{N}\frac{{\left| {T_i^ \odot \left( t \right)} \right|}}{{\left| {{T_i}\left( t \right)} \right|}}\\
P\left( {{X_i}\left( {t + 1} \right) =  - } \right) = 0
\end{array} \right.
\end{equation}
If $X_i(t) = -$, then 
\begin{equation}
\left\{ \begin{array}{l}
P\left( {{X_i}\left( {t + 1} \right) =  + } \right) = 0\\
P\left( {{X_i}\left( {t + 1} \right) =  \odot } \right) = \frac{\phi }{N}\frac{{\left| {T_i^ \odot \left( t \right)} \right|}}{{\left| {{T_i}\left( t \right)} \right|}}\\
P\left( {{X_i}\left( {t + 1} \right) =  - } \right) = \frac{\phi }{N}\frac{{\left| {T_i^ + \left( t \right)} \right| + \left| {T_i^ - \left( t \right)} \right|}}{{\left| {{T_i}\left( t \right)} \right|}}
\end{array} \right.
\end{equation}
If the agent $i$ holds opinion $\odot$, then the agent $i$ learns one of three opinions. That is, if $X_i(t) = \odot$, then 
\begin{equation}
P\left( {{X_i}\left( {t + 1} \right) =  m } \right) = \frac{\phi}{N} \frac{{\left| {T_i^ m (t) } \right|}}{{\left| {{T_i(t)}} \right|}}, {\rm where}\; m \in P .
\end{equation}
$\left| * \right|$ is the cardinal number of set $*$. In this model, it is similar to voter model. But agents only learn the teacher with similar opinions, mirroring the confidence bounds in Hegselmann-Krause model \cite{H_K_model_01,H_K_model_02, H_K_model_03}. If $T_i(t)$ is empty, then agent $i$ retains the current opinion. 

\subsection{Linking dynamics}
Linking dynamics occurs with probability $1 - \phi $ (see Fig. \hyperref[opinion_and_linking_bi_polarization]{2}). A directed link $(i,j) \in \mathcal E(t)$ is selected randomly and broken by $i$ or $j$ with pre-defined time-invariant breaking probability ${k_{[(i,j)]}}$. Then agent $i$ or $j$ rewires to a new agent $k$, who is neither in the current teacher set nor in the  student set of agent $i$ and $j$. Thus, the number of directed links is constant over time. We have
\begin{equation}
\!\!\!\!\!\!\begin{array}{*{20}{l}}
\begin{array}{l}
P[{A_{ij}}\left( {t + 1} \right) = 0,{A_{ik}}\left( {t + 1} \right) = 1\left| {{A_{ij}}\left( t \right) = 1,{A_{ik}}\left( t \right) = 0} \right.]\\
= \frac{{1 - \phi }}{{\left| {\mathcal E(t)} \right|}}\frac{{{k_{[(i,j)]}}}}{2}\frac{1}{{N - \left| {{T_i}(t)} \right| - \left| {{S_i}(t)} \right|}}
\end{array},
\end{array}
\end{equation}
and
\begin{equation}
\!\!\!\!\!\!\begin{array}{*{20}{l}}
\begin{array}{l}
P[{A_{ij}}\left( {t + 1} \right) = 0,{A_{kj}}\left( {t + 1} \right) = 1\left| {{A_{ij}}\left( t \right) = 1,{A_{kj}}\left( t \right) = 0} \right.]\\
= \frac{{1 - \phi }}{{\left| {\mathcal E(t)} \right|}}\frac{{{k_{[(i,j)]}}}}{2}\frac{1}{{N - \left| {{T_j}(t)} \right| - \left| {{S_j}(t)} \right|}}
\end{array}.
\end{array}
\end{equation}
We denote three link sets: ${E_0} = \{ {\overrightarrow { -  - } ,\overrightarrow { \odot  \odot } ,\overrightarrow { +  + } } \}$, ${E_1} = \{ {\overrightarrow { \odot  - } ,\overrightarrow { -  \odot } ,\overrightarrow { +  \odot } ,\overrightarrow { \odot  + } } \}$ and ${E_2} = \{ {\overrightarrow { +  - } ,\overrightarrow { -  + } } \}$. The subscript of $E$ refers to the distance between the two agents' opinions. If $[(i,j)]  \in {E_r}$, then ${k_{[(i,j)]}} = {K_r}$, where $K_r \in (0,1)$ and $r = \left\{ {0,1,2} \right\}$. 

\begin{definition}
	Weak homophily refers to ${K_0} < {K_1}$. Strong homophily refers to ${K_0} < {K_1}<{K_2}$.
\end{definition}

\section{An emergent game predicts the stability of bi-polarization}
\label{an_emergent_game}
Both opinion dynamics and linking dynamics are stochastic. And the size of the state space is too large to have its mean-field equations. In this section, for infinitely large system size, we first get its mean-field equations via time scale separation, and then analyze the stability of the polarization.

\begin{lemma} \cite{CCC_2024_wang}
	If $\phi  \to {0^ + }$, $\frac{L}{N} \to {0^ + }$ and ${x_ + }{x_ \odot }{x_ - } \ne 0$, then $\forall (i,j) \in \mathcal E(t)$, 
	\begin{equation}
	\lim_{t\to \infty}{\pi _{[(i,j)]}(t)} = {{\mathcal N}^ * }\left\{ \begin{array}{l}
	{{{x_{X_i}}{x_{X_j}}} \mathord{\left/
			{\vphantom {{{x_{X_i}}{x_{X_j}}} {{K_0}}}} \right.
			\kern-\nulldelimiterspace} {{K_0}}}\;\;[(i,j)]  \in {E_0}\\
	{{{x_{\tiny {X_i}}}{x_{X_j}}} \mathord{\left/
			{\vphantom {{{x_{X_i}}{x_{X_j}}} {{K_1}}}} \right.
			\kern-\nulldelimiterspace} {{K_1}}}\;\;[(i,j)]  \in {E_1}\\
	{{{x_{X_i}}{x_{X_j}}} \mathord{\left/
			{\vphantom {{{x_{X_i}}{x_{X_j}}} {{K_2}}}} \right.
			\kern-\nulldelimiterspace} {{K_2}}}\;\;[(i,j)]  \in {E_2}
	\end{array} \right.,
	\end{equation}
where ${{\mathcal N}^ * >0}$ is a normalization factor. 
\end{lemma}

\begin{lemma} \cite{CCC_2024_wang}
	If $\phi  \to {0^ + }$, $\frac{L}{N} \to {0^ + }$, $N \to +\infty $ and ${x_ + }{x_ \odot }{x_ - } \ne 0$, then the deterministic evolution of the fraction of opinion $+$ and opinion $-$ on the directed evolving graphs are given by 
	\begin{equation}
	\label{mean_field_equation}
	\begin{array}{l}
	{{\dot x}_ + } = \frac{{{x_ + }\left( {1 - {x_ + } - {x_ - }} \right)}}{{{K_1}\left( {\frac{{{x_ + }}}{{{K_1}}} + \frac{{1 - {x_ + } - {x_ - }}}{{{K_0}}} + \frac{{{x_ - }}}{{{K_1}}}} \right)}} - \frac{{{x_ + }\left( {1 - {x_ + } - {x_ - }} \right)}}{{{K_1}\left( {\frac{{{x_ + }}}{{{K_0}}} + \frac{{1 - {x_ + } - {x_ - }}}{{{K_1}}} + \frac{{{x_ - }}}{{{K_2}}}} \right)}}\\
	{{\dot x}_ - } = \frac{{\left( {1 - {x_ + } - {x_ - }} \right){x_ - }}}{{{K_1}\left( {\frac{{{x_ + }}}{{{K_1}}} + \frac{{1 - {x_ + } - {x_ - }}}{{{K_0}}} + \frac{{{x_ - }}}{{{K_1}}}} \right)}} - \frac{{\left( {1 - {x_ + } - {x_ - }} \right){x_ - }}}{{{K_1}\left( {\frac{{{x_ + }}}{{{K_2}}} + \frac{{1 - {x_ + } - {x_ - }}}{{{K_1}}} + \frac{{{x_ - }}}{{{K_0}}}} \right)}}
	\end{array}.
	\end{equation}
\end{lemma}
There are two trivial equilibrium sets, namely neutral consensus $(x_\odot = 1)$ and polarization $(x_\odot = 0)$. For polarization, we call  that agents reach radicalization if $x_+ = 1$ or $x_-=1$, and agents reach bi-polarization if $x_+ x_- \neq 0$.

\begin{definition}
	The bi-polarization is stable if the equilibrium set $\left( {x_ + ^*,x_ \odot ^*,x_ - ^*} \right) = \left( {\xi ,0,1 - \xi } \right)$ $(0 < \xi < 1)$ of Eq. \eqref{mean_field_equation} is asymptotically stable. 
\end{definition}

\begin{lemma} 
	\label{necessary_and_sufficient}
	Given weak homophily $({K_0} < {K_1})$, if $\phi  \to {0^ + }$, $\frac{L}{N} \to {0^ + }$, $N \to +\infty $ and ${x_ + }{x_ \odot }{x_ - } \ne 0$, then bi-polarization is stable if and only if $\frac{1}{{{K_1}}} - \frac{1}{{{K_2}}} < \frac{1}{{{K_0}}} - \frac{1}{{{K_1}}}$.
\end{lemma}
The proof of Lemma \ref{necessary_and_sufficient} is shown in Appendix \ref{Appendix_A}. Compared with \cite{CCC_2024_wang}, we prove that strong homophily is not necessary for stable bi-polarization. Besides,
we show that bi-polarization is unstable,
provided that $\frac{1}{{{K_1}}} - \frac{1}{{{K_2}}} = \frac{1}{{{K_0}}} - \frac{1}{{{K_1}}}$.

\begin{lemma}
	If $\phi  \to {0^ + }$, $\frac{L}{N} \to {0^ + }$ and ${x_ + }{x_ \odot }{x_ - } \ne 0$, denote $f_m$ as the average duration time between an agent with opinion $m$ and her teachers, where $m\in P$,
	then $f_m$ is the payoff of strategy $m$ in the following matrix game
	\begin{equation}
	\label{M}
	M = \begin{array}{*{20}{c}}&{\!\!\!\begin{array}{*{20}{c}}
		{ + \;}& \odot &{\;\, - }
		\end{array}}\\
	{\begin{array}{*{20}{c}}
		+ \\
		\odot \\
		- 
		\end{array}}&{\!\!\!\left( {\begin{array}{*{20}{c}}
			{\frac{1}{{{K_0}}}}&{\frac{1}{{{K_1}}}}&{\frac{1}{{{K_2}}}}\\
			{\frac{1}{{{K_1}}}}&{\frac{1}{{{K_0}}}}&{\frac{1}{{{K_1}}}}\\
			{\frac{1}{{{K_2}}}}&{\frac{1}{{{K_1}}}}&{\frac{1}{{{K_0}}}}
			\end{array}} \right)}
	\end{array}.
	\end{equation}
\end{lemma}

\noindent \textbf{Proof.} Denote ${D_{[(i,j)]}}$ as the duration time of the directed link $[(i,j)] \in E_r$, where $r=\{0,1,2\}$. $P\left( {{D_{[(i,j)]}} = n} \right) = {\left( {1 - {K_r}} \right)^{n - 1}}{K_r},\left( {n = 1,2, \cdots } \right)$. That is, ${D_{[(i,j)]}}$ follows the geometric distribution with successful probability $K_r$ \cite{Wu_Bridging,Du_CPB_2022}. $1/{K_r}$ is thus the average duration time of directed link $[(i,j)]$. Hence, the average duration time between an agent with opinion $+$ and her teachers is $f_+ = \frac{{{x_ + }}}{{{K_0}}} + \frac{{{x_ \odot }}}{{{K_1}}} + \frac{{{x_ - }}}{{{K_2}}}$. Similarly, ${f_ \odot } = \frac{{{x_ + }}}{{{K_1}}} + \frac{{{x_ \odot }}}{{{K_0}}} + \frac{{{x_ - }}}{{{K_1}}}$ and ${f_ - } = \frac{{{x_ + }}}{{{K_2}}} + \frac{{{x_ \odot }}}{{{K_1}}} + \frac{{{x_ - }}}{{{K_0}}}$. 
$f_m$ can be regarded as the payoff of strategy $m$ of the payoff matrix $M$, where $m\in P$. \hfill$\square$

\begin{remark}
	For an agent with opinion $m_1$, the fraction of her teachers who adopt opinion $m_2$ is $ \frac{{{\pi _{\tiny\overrightarrow {{m_1}{m_2}} }}}}{{{\pi _{\tiny\overrightarrow {{m_1} + } }} + {\pi _{\tiny\overrightarrow {{m_1} \odot } }} + {\pi _{\tiny\overrightarrow {{m_1} - } }}}} = \frac{{{x_{{m_2}}}/{K_r}}}{{{f_{{m_1}}}}}$, where $f_{{m_1}}$ is the payoff of strategy $m_1$ of the emergent game $M$, $m_1,m_2 \in P, \overrightarrow {m_1 m_2 }\in E_r$ and $r \in \{0,1,2\}$.
	Therefore the emergent game also captures the likelihood that a focal individual finds a teacher with a certain opinion. This is crucial for opinion evolution when opinions are about to update. 
\end{remark}

\begin{lemma}
	\label{nash_equilibrium}
	Given weak homophily, the emergent game $M$ has an internal Nash equilibrium if and only if $\frac{1}{{{K_1}}} - \frac{1}{{{K_2}}} < \frac{1}{{{K_0}}} - \frac{1}{{{K_1}}}$.
\end{lemma}

\noindent \textbf{Proof.} $\,$
The replicator equation of the emergent game $M$ is ${\dot x_s} = {x_s}( {{f_s} - \bar f} )$ with $x_+ + x_\odot + x_- = 1$, where $s \in \{+,-\}$, $\bar{f} = {x_ + }{f_ + } + {x_ \odot }{f_ \odot } + {x_ - }{f_ - }$. Let ${\dot x_s} = 0$, given ${K_0} < {K_1}$, the unique internal fixed point $\left( \frac{{{K_2}\left( {{K_1} - {K_0}} \right)}}{{{K_0}{K_1} - 4{K_0}{K_2} + 3{K_1}{K_2}}},\frac{{{K_0}{K_1} - 2{K_0}{K_2} + {K_1}{K_2}}}{{{K_0}{K_1} - 4{K_0}{K_2} + 3{K_1}{K_2}}},\right.$\\$\left. \frac{{{K_2}\left( {{K_1} - {K_0}} \right)}}{{{K_0}{K_1} - 4{K_0}{K_2} + 3{K_1}{K_2}}} \right)$ exists if and only if $\frac{1}{{{K_1}}} - \frac{1}{{{K_2}}} < \frac{1}{{{K_0}}} - \frac{1}{{{K_1}}}$. And this internal fixed point of the replicator equation is an internal Nash equilibrium of the emergent game $M$ \cite{Hofbauer_1998}. \hfill$\square$

\begin{remark}
	Nash equilibrium refers to a state in which no player obtains more if one player unilaterally changes her strategy \cite{ESS_taylor, Nash_equilibrium}. That is, the average duration time of an agent does not increase if she unilaterally changes her opinion in our emergent game $M$.
\end{remark}

\begin{theorem}
	\label{theorem_1}
	Given weak homophily, if $\phi  \to {0^ + }$, $\frac{L}{N} \to {0^ + }$, $N \to +\infty $ and ${x_ + }{x_ \odot }{x_ - } \ne 0$, then the bi-polarization is stable if and only if the emergent game $M$ has an internal Nash equilibrium, i.e., $\frac{1}{{{K_1}}} - \frac{1}{{{K_2}}} < \frac{1}{{{K_0}}} - \frac{1}{{{K_1}}}$.
\end{theorem}
Proof of Theorem \ref{theorem_1} follows directly from Lemma \ref{necessary_and_sufficient} and Lemma \ref{nash_equilibrium}.

What does $\frac{1}{{{K_1}}} - \frac{1}{{{K_2}}} < \frac{1}{{{K_0}}} - \frac{1}{{{K_1}}}$ imply?
To shed game light on opinion formation, we give the following definitions on active payoff $f^{\rm{act}}_m$ and inactive payoff $f^{\rm{inact}}_m$, where $m \in P$.

\begin{figure}
	\centering
	\includegraphics[scale=0.50]{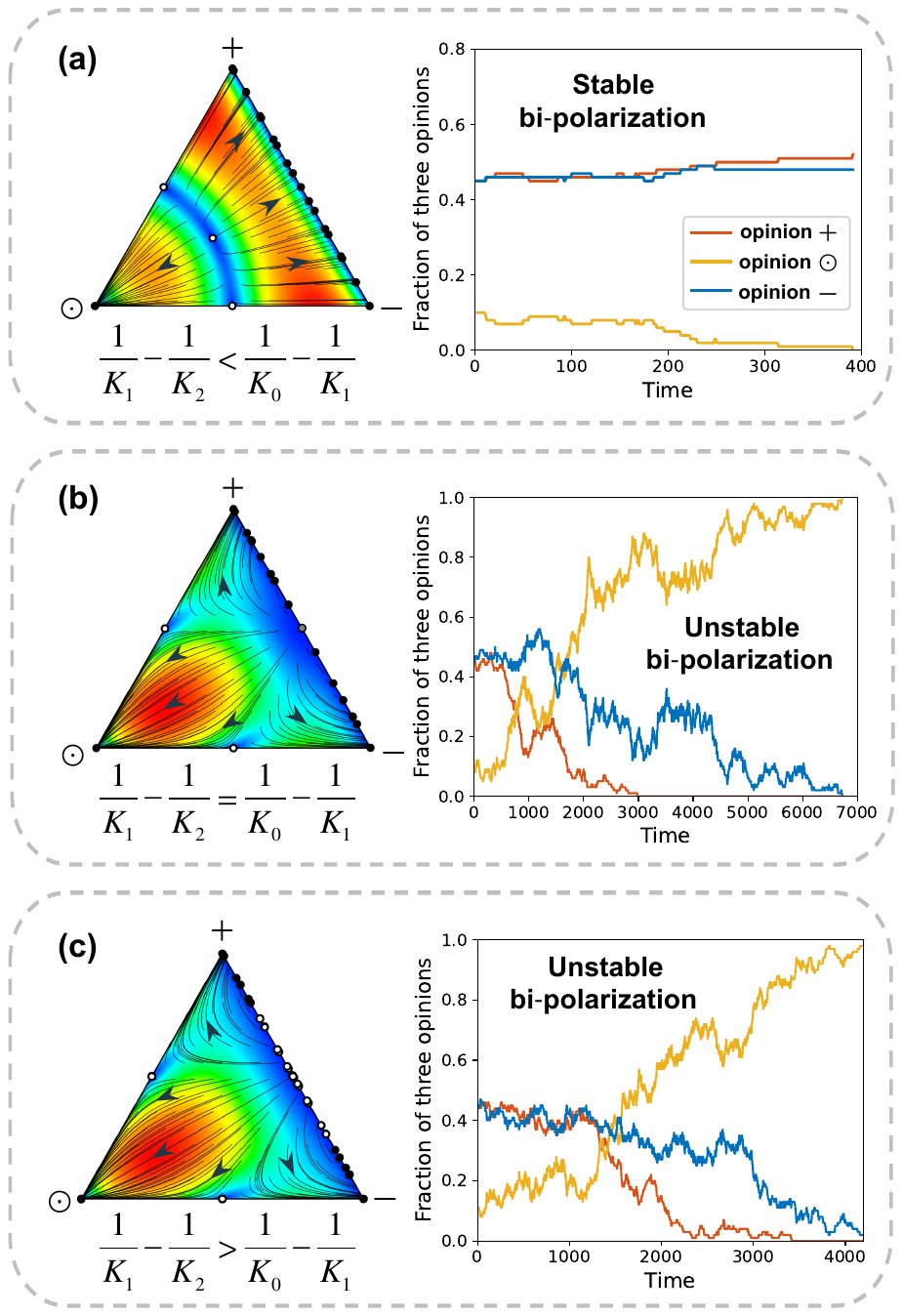}
	\caption{\textbf{Opinion formation of conservative opinion $(+)$, liberal opinion $(-)$, and neutral opinion $( \odot )$.} \textbf{(a)} Given $K_0 < K_1$, if $\frac{1}{{{K_1}}} - \frac{1}{{{K_2}}} < \frac{1}{{{K_0}}} - \frac{1}{{{K_1}}}$, then the bi-polarization is stable. Parameters: $K_0=0.2, K_1=0.6,K_2=0.3$. \textbf{(b)} Given $K_0 < K_1$, if $\frac{1}{{{K_1}}} - \frac{1}{{{K_2}}} = \frac{1}{{{K_0}}} - \frac{1}{{{K_1}}}$, then the bi-polarization is unstable. Parameters: $K_0=0.2, K_1=0.3,K_2=0.6$. \textbf{(c)} Given $K_0 < K_1$, if $\frac{1}{{{K_1}}} - \frac{1}{{{K_2}}} > \frac{1}{{{K_0}}} - \frac{1}{{{K_1}}}$, then the bi-polarization is unstable. Parameters: $K_0=0.2, K_1=0.3,K_2=0.9$. }
	\label{stability}
\end{figure}

\begin{definition}
	$f_m = f^{\rm{act}}_m + f^{\rm{inact}}_m$, where $m \in P$. $f_ + ^{{\rm{act}}} = f_ - ^{{\rm{act}}} = \frac{{{x_ \odot }}}{{{K_1}}},f_ \odot ^{{\rm{act}}} = \frac{{{x_ + } + {x_ - }}}{{{K_1}}},f_ + ^{{\rm{inact}}} = \frac{{{x_ + }}}{{{K_0}}} + \frac{{{x_ - }}}{{{K_2}}},f_ - ^{{\rm{inact}}} = \frac{{{x_ - }}}{{{K_0}}} + \frac{{{x_ + }}}{{{K_2}}},f_ \odot ^{{\rm{inact}}} = \frac{{{x_ \odot }}}{{{K_0}}}$.
	$E_1$ is an active link set. Active links, leading to active payoff $f^{\rm{act}}_m$, i.e., those links that can lead to imitation. $E_0$ and $E_2$ are inactive link sets. Inactive links, leading to inactive payoff $f^{\rm{inact}}_m$, i.e., those links that cannot lead to imitation. 
\end{definition}

We provide two game perspectives of the necessary and sufficient condition in Theorem \ref{theorem_1}. On the one hand, we focus on risk dominance. Let us consider a population, in which all the opinions are equally abundant, i.e., $({x_ + }, {x_ \odot }, {x_ - }) = \left(\frac{1}{3},\frac{1}{3},\frac{1}{3}\right)$.
The inactive payoff of extreme opinions is $ \frac{2}{3K_0} + \frac{2}{3K_2}$. The active payoff of all opinions is $\frac{4}{3K_1}$. If the active payoff of all opinions is smaller than the inactive payoff of extreme opinions, i.e., $\frac{1}{{{K_1}}} - \frac{1}{{{K_2}}} < \frac{1}{{{K_0}}} - \frac{1}{{{K_1}}}$, then the bi-polarization is stable. On the other hand, we provide a game theoretical explanation based on evolutionary stable strategy. We address the increment payoff induced by a small fraction of neutral opinions on the inactive and active payoff, respectively. Denote the fraction of opinion $\odot$ as $\varepsilon $. The average inactive payoff of extreme opinions is $\pi_1 = \frac{x_+}{{1 - \varepsilon }}\left( {\frac{x_+}{{{K_0}}} + \frac{x_-}{{{K_2}}}} \right) + \frac{{x_-}}{{1 - \varepsilon }}\left( {\frac{x_+}{{{K_2}}} + \frac{x_-}{{{K_0}}}} \right)$. The average active payoff of extreme opinions is $\pi_2 = \frac{x_+}{{1 - \varepsilon }}\frac{\varepsilon}{{{K_1}}}  + \frac{{x_-}}{{1 - \varepsilon }}\frac{\varepsilon}{{{K_1}}} = \frac{\varepsilon }{{{K_1}}}$. Let $\varepsilon  \to {0^ + }$, we have that ${\left. {\frac{{\partial {\pi_1}}}{{\partial \varepsilon }}} \right|_{\varepsilon  = 0}} =  - \frac{{2{K_0}{x_+^2} + {K_2}\left( {1 - 2{x_+^2}} \right)}}{{{K_0}{K_2}}}$ and ${\left. {\frac{{\partial {\pi_2}}}{{\partial \varepsilon }}} \right|_{\varepsilon  = 0}} = \frac{1}{{{K_1}}}$. Let us consider a population, in which extreme opinions, are equally abundant, i.e., $x_+ = x_- = \frac{1}{2}$. If the rate of decreasing in inactive payoff is larger than that of increasing in active payoff, i.e. $\frac{1}{2}\left( {\frac{1}{{{K_0}}} + \frac{1}{{{K_2}}}} \right) > \frac{1}{{{K_1}}}$, which is the necessary and sufficient condition, then bi-polarization is stable.

To shed further game theoretical insight, we mirror the opinion formation in our model to coordination games.
\begin{lemma} 
	\label{neutral_consensus}
	Given weak homophily $({K_0} < {K_1})$, if $\phi  \to {0^ + }$, $\frac{L}{N} \to {0^ + }$, $N \to +\infty $ and ${x_ + }{x_ \odot }{x_ - } \ne 0$, then neutral consensus is asymptotically stable.
\end{lemma}

\begin{lemma}
	\label{interal_equilibrium}
	Given weak homophily $({K_0} < {K_1})$, there is an unique unstable internal equilibrium of Eq. \eqref{mean_field_equation} if and only if $\frac{1}{{{K_1}}} - \frac{1}{{{K_2}}} < \frac{1}{{{K_0}}} - \frac{1}{{{K_1}}}$.
\end{lemma}

Based on Lemma \ref{necessary_and_sufficient} and Lemma \ref{neutral_consensus},
both bi-polarization and neutral consensus are stable, given weak homophily with $\frac{1}{{{K_1}}} - \frac{1}{{{K_2}}} < \frac{1}{{{K_0}}} - \frac{1}{{{K_1}}}$.
Lemma \ref{interal_equilibrium} shows that there is at most one internal equilibrium.
And the internal equilibrium is unstable if there is any.
Altogether, the dynamics of opinion formation mirror a coordination game. 
A classical coordination game has two stable fixed points (evolutionary stable strategies) and a unique unstable fixed point (unstable Nash equilibrium). 
Therefore, the opinion transition from neutral opinion to bi-polarization mirrors the convention formation in coordination games \cite{Thomas_micr_macr}. In the classical coordination game, the unique internal fixed point captures the attraction basin. The closer the internal fixed point is to a strategy, the smaller the attraction basin of that strategy is. Inspired by this, this game perspective allows us to make use of the unstable internal equilibrium $(x_{\rm{internal\,+}}^*, x_{\rm{internal\,\odot}}^*, x_{\rm{internal\,-}}^*)$ of mean-field equation Eq.
\eqref{mean_field_equation} to measure the attraction basin as in coordination games.
We can mirror the attraction basin of neutral consensus as  
$(x_{\rm{internal\,\odot}}^*, 1)$, where $1$ refers to the neutral consensus in coordination games, although this is not exactly true. 
Thus large $x_{\rm{internal\,\odot}}^*$ implies the expansion of the attraction basin of bi-polarization.
Note-worthily, we find that 
$x_{\rm{internal\,+}}^*+x_{\rm{internal\,-}}^*$ increases if and only if $x_{\rm{internal\,+}}^*$ increases. This is true because the model is symmetric for both extreme opinions.
It also implies that there is no zero-sum game between opinion $+$ and opinion $-$. Furthermore, we show that the emergent game $M$ suffices to predict the attraction basin expansion of bi-polarization.
For the emergent game $M$, denote $f_{\rm{ratio}}$ as the ratio of the active payoff of three opinions to the inactive payoff of three opinions, i.e., $f_{\rm{ratio}} = \frac{{f_ + ^{{\rm{act}}} + f_ \odot ^{{\rm{act}}} + f_ - ^{{\rm{act}}}}}{{f_ + ^{{\rm{inact}}} + f_ \odot ^{{\rm{inact}}} + f_ - ^{{\rm{inact}}}}} = \frac{(1+x_\odot)/K_1}{1/K_0+(1-x_\odot)/K_2}$. 
If $K_0$ or $K_2$ decreases or $K_1$ increases, then $f_{\rm{ratio}}$ decreases.
On the other hand,
if $K_0$ or $K_2$ decreases or $K_1$ increases,
then $x_{\rm{internal\,\odot}}^*$ increases and the attraction basin of bi-polarization is larger.
In other words, 
the attraction basin for stable bi-polarization becomes larger,
provided that $f_{\rm{ratio}}$ decreases.
Therefore active payoff and inactive payoff provide a universal explanation for how breaking probabilities alters the attraction basin of polarization. 

\section{Simulations}
\label{simulations}
We perform Monte Carlo simulations to verify our theoretical results. For each of the following simulations, we consider $N=100$, $L=4$ and $\phi = 0.01$. Firstly, we validate Lemma \ref{necessary_and_sufficient}. We assume the initial fraction of three opinions is $\left( {x_ + ^{{\rm{initial}}},x_ \odot ^{{\rm{initial}}},x_ - ^{{\rm{initial}}}} \right) = \left( {0.45,0.1,0.45} \right)$. Given $K_0 < K_1$, we set $K_0=0.2, K_1=0.6,K_2=0.3$, which satisfies $\frac{1}{{{K_1}}} - \frac{1}{{{K_2}}} < \frac{1}{{{K_0}}} - \frac{1}{{{K_1}}}$. Opinion $\odot$ goes extinction. Opinion $+$ and opinion $-$ exist stably. The bi-polarization is stable, as shown in Fig. \hyperref[stability]{3(a)}. In this case, homophily is weak. Therefore, strong homophily is not necessary for stable bi-polarization. We set $K_0=0.2,K_1=0.3,K_2=0.6$, which satisfies $\frac{1}{{{K_1}}} - \frac{1}{{{K_2}}} = \frac{1}{{{K_0}}} - \frac{1}{{{K_1}}}$. The equilibrium set $\left( {x_ + ^*,x_ \odot ^*,x_ - ^*} \right) = \left( {\xi ,0,1 - \xi } \right)$ is not all stable. The population reaches consensus on opinion $\odot$. Opinion $+$ and opinion $-$ are extinct. Thus the bi-polarization is unstable as shown in see Fig. \hyperref[stability]{3(b)}. We set $K_0=0.2, K_1=0.3,K_2=0.9$, which satisfies $\frac{1}{{{K_1}}} - \frac{1}{{{K_2}}} > \frac{1}{{{K_0}}} - \frac{1}{{{K_1}}}$. And the bi-polarization is unstable as shown in see \hyperref[stability]{3(c)}. Phase diagrams are drawn by Dynamo \cite{Dynamo}.

Secondly, we validate $f_{\rm{ratio}}$ as a game indicator for the attraction basin of stable bi-polarization. As shown in Fig. \hyperref[stability]{4}, if $K_0$ or $K_2$ decreases, or $K_1$ increases, the attraction basin for stable bi-polarization increases in measure. Simultaneously, $f_{\rm{ratio}}$ decreases if $K_0$ or $K_2$ decreases or $K_1$ increases. Phase diagrams are the theoretical results and pie charts are the simulation results which show the size of the attraction basin. We set the initial state of each round is random and run 1000 rounds. 

\begin{figure}
	\centering
	\includegraphics[scale=0.25]{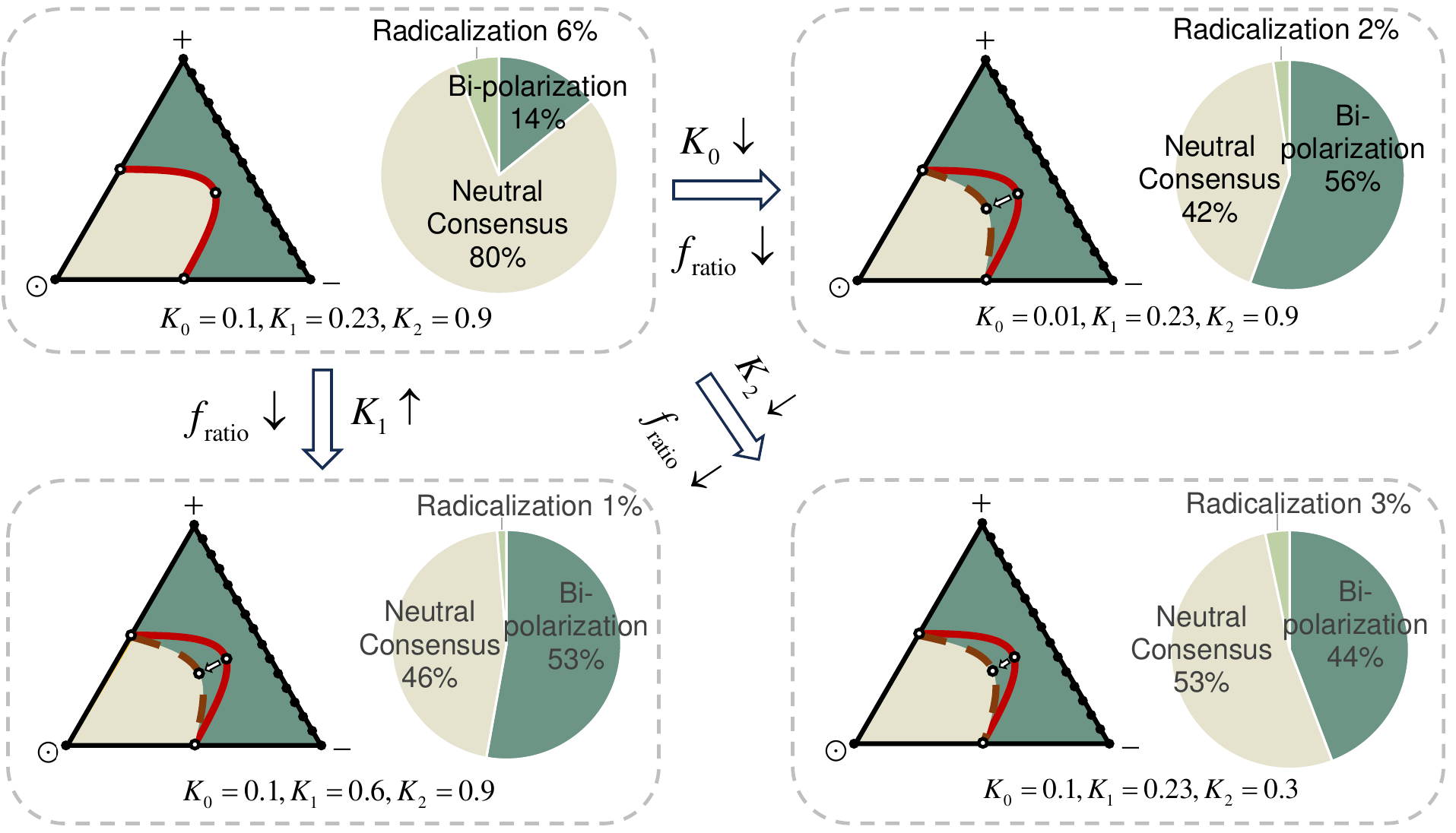}
	\caption{$f_{\rm{ratio}}$ is a game indicator for the attraction basin of stable bi-polarization. }
	\label{attraction_basin}
\end{figure}

\section{Empirical validation}
\label{empirical_validation}
\begin{figure}
	\centering
	\includegraphics[scale=0.28]{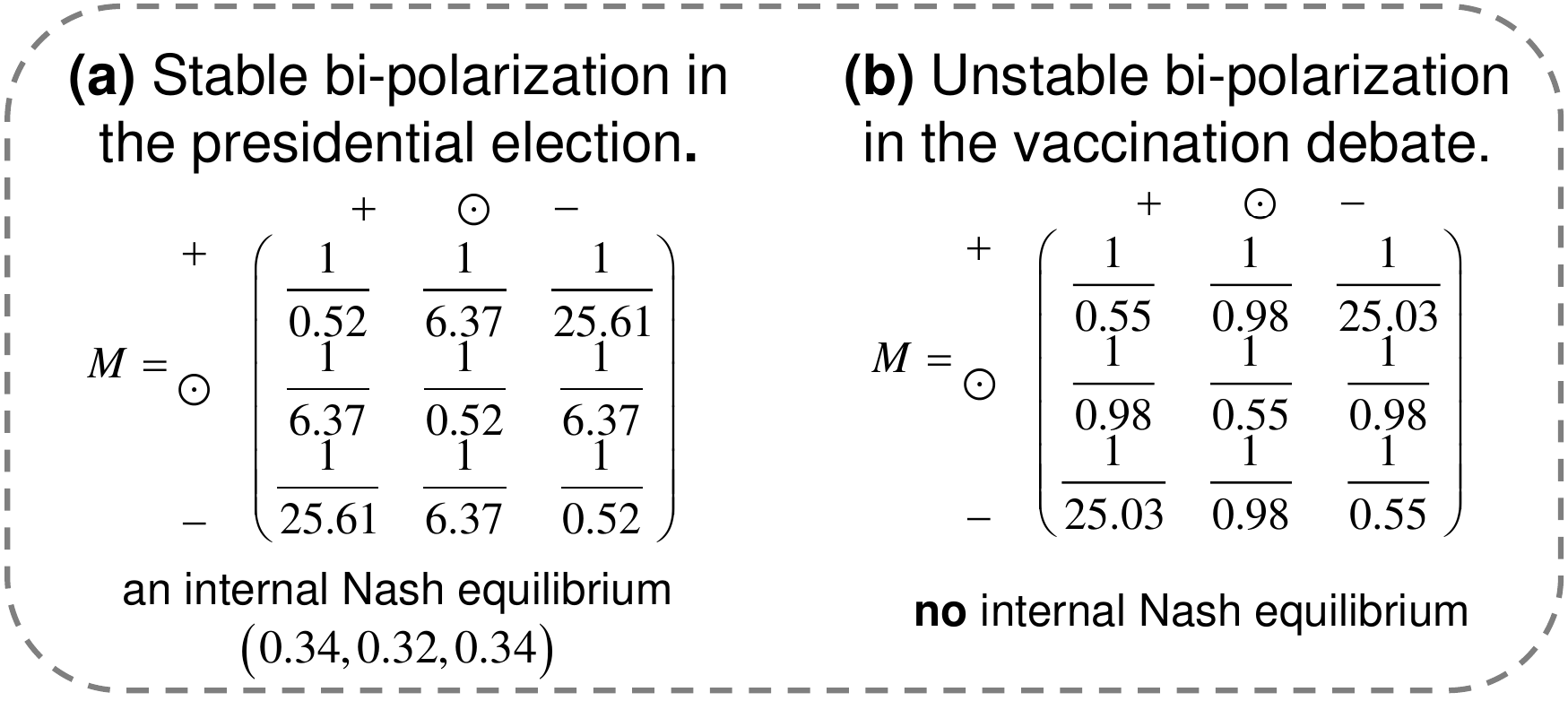}
	\caption{This game perspective facilitates us to reveal the stability of bi-polarization in the empirical system.} 
	\label{validate_data}
\end{figure}

Wang et al. collected the retweets of the first, second and third presidential debates in the 2016 U.S. presidential election \cite{PRX_echo_chamber}. There are Pro-Trump supporters $(+)$, Pro-Hillary supporters $(-)$ and Neutral $(\odot)$. We assume that the online social graph has reached the stationary regime at the third debate. The fraction of three opinions is $({x_ + }, {x_ \odot }, {x_ - }) = (0.583, 0.004, 0.413)$ at the third debate based on the data. Based on our model, we have the best fitting for breaking probabilities $ \left( {{K_0 ^ *},{K_1 ^ *},{K_2 ^ *}} \right) \approx {{\mathcal N}^ * }\left( {0.52,6.37,25.61} \right)$. There is no unique solution for the three breaking probabilities due to the positive normalization factor ${{\mathcal N}^ * >0}$, but it is adequate to capture the order. The emergent game $M$ has an internal Nash equilibrium, which explains why the bi-polarization is stable by Theorem 1 (see Fig. \hyperref[validate_data]{5(a)}). From the empirical data, opinion $+$ and opinion $-$ coexist stably. Opinion $\odot$ are absorbed by either opinion $+$ or opinion $-$ (see Fig. \hyperref[stable_bipolarization]{1(a)}). More details can be found in Appendix \ref{Appendix_B}.

Lachi et al. provided the dataset of Italian retweets on the vaccination debate between 2019 and 2022 \cite{Vaccine_Lachi_twitter}. Based on the number of tweets retweeted by agents, we divide agents into three types, i.e., No-Vaccine supporters $(+)$, Pro-Vaccine supporters $(-)$ and Neutral $(\odot)$. The number of tweets on the vaccination debate began to spike in August 2020 (see the Fig. 9 in \cite{Vaccine_Lachi_twitter}). It suggests that the social online graph has reached the stationary regime that time. The fraction of three opinions is $({x_ + }, {x_ \odot }, {x_ - }) = (0.356,0.005,0.639)$ in August 2020. Similarly, we have the best fitting for breaking probabilities $\left( {{K_0 ^ *},{K_1 ^ *},{K_2 ^ *}} \right) \approx {{\mathcal N}^ * }\left( {0.55,0.98,25.03} \right)$. However, the emergent game $M$ does \emph{not} have an internal Nash equilibrium, which explains why the bi-polarization is not stable by Theorem 1 (see Fig. \hyperref[validate_data]{5(b)}). From the empirical data, the fraction of the three opinions vary considerably and opinion $+$ and opinion $-$ do not coexist (see Fig. \hyperref[stable_bipolarization]{1(b)}). More details can be found in Appendix \ref{Appendix_B}.

\section{Conclusions}
We focus on discrete opinions whereas previous works focus on the continuous opinions \cite{Automatica_polarization,PRL_Modeling}. In fact, discrete opinions are more ubiquitous in the real world. For example, political parties are discrete in number. We adopt learning the similar for opinion adjustment, which is similar to the confidence bounds in Hegselmann-Krause model \cite{H_K_model_01, Automatica_H_K_2024}. It is challenging to find the necessary and sufficient condition for polarization in Hegselmann-Krause model \cite{H_K_model_02,H_K_model_03}. Our result thus, to the best of our knowledge, gives the first necessary and sufficient condition for a discrete alternative.

Compared with our previous work \cite{CCC_2024_wang}, our contributions are four-fold:
i) strong homophily is not necessary for stable bi-polarization in this brief;
ii) in \cite{CCC_2024_wang}, the necessary and sufficient condition of stable bi-polarization is not complete.
Instead,
here we give the necessary and sufficient condition by showing that the bi-polarization is unstable if $\frac{1}{{{K_1}}} - \frac{1}{{{K_2}}} = \frac{1}{{{K_0}}} - \frac{1}{{{K_1}}}$;
iii) we establish a bridge between the necessary and sufficient condition of stable bi-polarization and game theory.
Our game perspective sheds light on the network structure of the system, leading to more elusive insights on opinion formation;
iv) the necessary and sufficient condition is validated by empirical data.   

Most previous works analyze the stability of bi-polarization on static graphs rather than dynamical graphs \cite{Dandekar_PNAS_2012,Automatica_polarization}, for an exception see \cite{Emergence_polarization_PRL}. In \cite{Emergence_polarization_PRL}, stable bi-polarization is predicted by fitting both linking dynamics and opinions updating parameters. Our model is able to capture the stability of bi-polarization by calculating three breaking probabilities of the linking dynamics alone. We do not take into account the opinion updating parameters. Our model is thus more likely to fulfill the principle of Occam's Razor \cite{Ockhams_Razor}. 

There has been increasing attention on the evolution of opinions from a game perspective in recent years. The utility function (game) is crucial in the sense that all the principles of updating are encoded in those utility functions \cite{opinion_game_TAC_2023, opinion_game_Chaos_2022, opinion_game_Fu_2018}. Agents in our model have \emph{no} payoff in mind when updating opinions. The emergent two-player three-strategy game predicts the stability of bi-polarization. Despite these potential benefits, our model has a limitation. The graph evolves much faster than the opinions \cite{PRL_Modeling, Emergence_polarization_PRL, Wu_Bridging, A_tale_of_two_games, Du_CPB_2022}. We attempt to seek general results in any time scale with the method in Ref. \cite{wang_opinion_time} in the future. 

To sum up, our work gives the necessary and sufficient condition of stable bi-polarization on dynamical directed graphs and predicts the stability of bi-polarization via an emergent game. 

\appendix
\renewcommand{\theequation}{A\arabic{equation}}
\section{Proof of Lemma 3.} \label{Appendix_A}
For the evolution of opinions, we denote $Q_{{x_m}}^ + \,(Q_{{x_m}}^ - )$ as the transition probability that opinion $s$ increases (decreases) by $1/N$, where $m \in \{+,-\}$. Based on the voter-like model, we have
\begin{equation}
\begin{array}{l}
\left\{ \begin{array}{l}
Q_{{x_ + }}^ +  = {x_ \odot }\frac{{{\pi _{\tiny\overrightarrow { \odot  + } }}}}{{{\pi _{\tiny\overrightarrow { \odot  + } }} + {\pi _{\tiny\overrightarrow { \odot  \odot } }} + {\pi _{\tiny\overrightarrow { \odot  - } }}}}\\
Q_{{x_ + }}^ -  = {x_ + }\frac{{{\pi _{\tiny\overrightarrow { +  \odot } }}}}{{{\pi _{\tiny\overrightarrow { +  + } }} + {\pi _{\tiny\overrightarrow { +  \odot } }} + {\pi _{\tiny\overrightarrow { +  - } }}}}
\end{array} \right.\\
\left\{ \begin{array}{l}
Q_{{x_ - }}^ +  = {x_ \odot }\frac{{{\pi _{\tiny\overrightarrow { \odot  - } }}}}{{{\pi _{\tiny\overrightarrow { \odot  - } }} + {\pi _{\tiny\overrightarrow { \odot  \odot } }} + {\pi _{\tiny\overrightarrow { \odot  + } }}}}\\
Q_{{x_ - }}^ -  = {x_ - }\frac{{{\pi _{\tiny\overrightarrow { -  \odot } }}}}{{{\pi _{\tiny\overrightarrow { -  \odot } }} + {\pi _{\tiny\overrightarrow { -  - } }} + {\pi _{\tiny\overrightarrow { -  + } }}}}
\end{array} \right.
\end{array}.
\end{equation}
The probability that all of opinions remain constant is $1 - Q_{{x_+}}^ +  - Q_{{x_+}}^ - -Q_{{x_-}}^ +  - Q_{{x_-}}^ -$. 
For large population size, i.e., $N \to +\infty $, the mean-field equations are given by 
\begin{equation}
\left\{ \begin{array}{l}
{{\dot x}_ + } = Q_{{x_ + }}^ +  - Q_{{x_ + }}^ - \\
{{\dot x}_ - } = Q_{{x_ - }}^ +  - Q_{{x_ - }}^ - 
\end{array} \right.,
\end{equation}
with $x_+ + x_\odot + x_- = 1$. 
Let ${{\dot x}_ + } = 0$ and ${{\dot x}_ - } = 0$ in Eq. (A2), we obtain all the equilibria. There is at most one equilibrium set $\mathcal C ^* \buildrel \Delta \over = ( {x_ + ^*,x_ \odot ^*,x_ - ^*} ) = ( {\xi ,0,1 - \xi } )$, where $0 \le \xi \le 1$ and four isolated equilibria $( {x_ + ^*,x_ \odot ^*,x_ - ^*} )$: (1) $\left( {1/2,1/2,0} \right)$, (2) $\left( {0,1/2,1/2} \right)$, (3) $\left( {0,1,0} \right)$. Given $K_0 < K_1$, if $\frac{1}{{{K_1}}} - \frac{1}{{{K_2}}} < \frac{1}{{{K_0}}} - \frac{1}{{{K_1}}}$, then $(x_{\rm{internal\,+}}^*, x_{\rm{internal\,\odot}}^*, x_{\rm{internal\,-}}^*) = \left( \frac{{{K_2}\left( {{K_1} - {K_0}} \right)}}{{{K_0}{K_1} - 4{K_0}{K_2} + 3{K_1}{K_2}}},\right.$\\$\left.\frac{{{K_0}{K_1} - 2{K_0}{K_2} + {K_1}{K_2}}}{{{K_0}{K_1} - 4{K_0}{K_2} + 3{K_1}{K_2}}}, \frac{{{K_2}\left( {{K_1} - {K_0}} \right)}}{{{K_0}{K_1} - 4{K_0}{K_2} + 3{K_1}{K_2}}} \right)$ exists.
We analyze the stability of the equilibrium set $\mathcal C ^*$ with the aid of the Jacobian matrix. The Jacobian matrix is calculated by the following equations:
\begin{equation}
J\left( {x_ + ^ * ,x_ - ^ * } \right) = \left( {\begin{array}{*{20}{c}}
	{{{\left. {\displaystyle\frac{{\partial {{\dot x}_ + }}}{{\partial {x_ + }}}} \right|}_{\left( {x_ + ^ * ,x_ - ^ * } \right)}}}&{{{\left. {\displaystyle\frac{{\partial {{\dot x}_ + }}}{{\partial {x_ - }}}} \right|}_{\left( {x_ + ^ * ,x_ - ^ * } \right)}}}\\
	{{{\left. {\displaystyle\frac{{\partial {{\dot x}_ - }}}{{\partial {x_ + }}}} \right|}_{\left( {x_ + ^ * ,x_ - ^ * } \right)}}}&{{{\left. {\displaystyle\frac{{\partial {{\dot x}_ - }}}{{\partial {x_ - }}}} \right|}_{\left( {x_ + ^ * ,x_ - ^ * } \right)}}}
	\end{array}} \right)
\end{equation}
The eigenvalues of the corresponding the Jacobian matrices of each equilibrium are shown as follows:

\noindent (a) $\left( {\xi ,0,1 - \xi } \right)$, where $0 \le \xi \le 1$: \\${\lambda _1} = 0$,${\lambda _2} = [{K_0}{K_2}( {{K_0} - {K_1}} ) + ( {{K_2} - {K_0}} )( {K_0}{K_1} + 2{K_0}{K_2} - {K_1}{K_2} )\xi  + ( {K_0} - {K_2} )( {{K_0}{K_1} + 2{K_0}{K_2} - {K_1}{K_2}} ){\xi ^2}]$\\$[ {\xi {K_1}{K_2} + ( {1 - \xi } ){K_0}{K_1}} ]^{ - 1}[ {\xi {K_0}{K_1} + ( {1 - \xi } ){K_1}{K_2}} ]^{ - 1}$.
\\Given $K_0 < K_1$, if $\frac{1}{{{K_1}}} - \frac{1}{{{K_2}}} < \frac{1}{{{K_0}}} - \frac{1}{{{K_1}}}$, then ${\lambda _2}<0$ and the equilibrium set $\mathcal C ^*$ is stable. There is stable bi-polarization. Given $K_0 < K_1$, if $\frac{1}{{{K_1}}} - \frac{1}{{{K_2}}} > \frac{1}{{{K_0}}} - \frac{1}{{{K_1}}}$, then some of the equilibrium set $\mathcal C ^*$ is stable but others are not. There is unstable bi-polarization. 
Given $K_0<K_1$, if $\frac{1}{{{K_1}}} - \frac{1}{{{K_2}}} = \frac{1}{{{K_0}}} - \frac{1}{{{K_1}}}$, then $K_1<K_2$. Let $x_+ = x_- = 1/2-\varepsilon$, where $\varepsilon>0$. Thus ${{\dot x}_ \odot } = -{{\dot x}_ + } -{{\dot x}_ - } = \frac{{4{\varepsilon^2}}}{{{K_1}}}\left( {\frac{1}{{{K_1}}} - \frac{1}{{{K_2}}}} \right) + \frac{{8{\varepsilon^3}}}{{{K_1}}}\left( {\frac{1}{{{K_2}}} - \frac{1}{{{K_1}}}} \right)$. If $\varepsilon  \to {0^ + }$, then ${\dot x_ \odot } \to {0^ + }$ and $x_ \odot$ increases. There is unstable bi-polarization. Therefore, given $K_0 < K_1$, $\frac{1}{{{K_1}}} - \frac{1}{{{K_2}}} < \frac{1}{{{K_0}}} - \frac{1}{{{K_1}}}$ is the necessary and sufficient condition for stable bi-polarization.

\noindent (b) $\left( {1/2,1/2,0} \right)$: \\${\lambda _1} = \frac{{2{K_0}\left( {{K_1} - {K_0}} \right)}}{{{{\left( {{K_0} + {K_1}} \right)}^2}}},{\lambda _2} = \frac{{{K_1}\left( {{K_0} - {K_2}} \right)}}{{\left( {{K_0} + {K_1}} \right)\left( {{K_1} + {K_2}} \right)}}$. 
\\Given $K_0 < K_1$, ${\lambda _1}>0$ and ${\lambda _2}<0$. Thus the equilibrium (1) is unstable.

\noindent (c) $\left( {0,1/2,1/2} \right)$: same as (b).

\noindent (d) $\left( {0,1,0} \right)$: \\${\lambda _1} = {\lambda _2} = \frac{{{K_0} - {K_1}}}{{{K_1}}} < 0$. Given $K_0 < K_1$, ${\lambda _1}={\lambda _2}<0$. Thus the equilibrium (3) is stable.

\noindent (e) $(x_{\rm{internal\,+}}^*, x_{\rm{internal\,\odot}}^*, x_{\rm{internal\,-}}^*)$: \\${\lambda _1} = \frac{{{K_0}{K_1}\left( {{K_1} - {K_0}} \right)\left( {{K_2} - {K_0}} \right)\left( {{K_0}{K_1} - 2{K_0}{K_2} + {K_1}{K_2}} \right)}}{{{{\left( {{K_0}K_1^2 - 2K_0^2{K_2} + K_1^2{K_2}} \right)}^2}}}$, \\${\lambda _2} = \frac{{{K_0}\left( {{K_1} - {K_0}} \right)\left( {{K_0}{K_1} - 4{K_0}{K_2} + 3{K_1}{K_2}} \right)\left( {{K_0}{K_1} - 2{K_0}{K_2} + {K_1}{K_2}} \right)}}{{{{\left( {{K_0}K_1^2 - 2K_0^2{K_2} + K_1^2{K_2}} \right)}^2}}}$. \\Given $K_0 < K_1$, if $\frac{1}{{{K_1}}} - \frac{1}{{{K_2}}} < \frac{1}{{{K_0}}} - \frac{1}{{{K_1}}}$, then ${\lambda _2}>0$. Thus the internal equilibrium is unstable.

\section{Data description.} \label{Appendix_B}
Wang et al. collected the retweet graphs (now ``X") in the final stages of the 2016 U.S. presidential election \cite{PRX_echo_chamber}. In the third debate, $N=25338$, $x_+ = 0.583$ (the fraction of Trump supports), $x_\odot = 0.004$ (the fraction of neutral supports) and $x_- = 0.413$ (the faction of Clinton supports). The fraction of nine directed links is $( {{\pi _{\scriptsize\overrightarrow { -  - } }},{\pi _{\scriptsize\overrightarrow { \odot  - } }},{\pi _{\scriptsize\overrightarrow { +  - } }},{\pi _{\scriptsize\overrightarrow { -  \odot } }},{\pi _{\scriptsize\overrightarrow { \odot  \odot } }},{\pi _{\scriptsize\overrightarrow { +  \odot } }},{\pi _{\scriptsize\overrightarrow { -  + } }},{\pi _{\scriptsize\overrightarrow { \odot  + } }},{\pi _{\scriptsize\overrightarrow { +  + } }}} ) = (0.32282,0.00033,0.00778,0.00046,0.00266,0.00018,\\0.01102,0.00028,0.65447)$. We do not take the self-loop and multiple links into account in this scenario. Based on the stationary distribution in our model, we calculate three breaking probabilities with best fitting $\left( {{K_0 ^ *},{K_1 ^ *},{K_2 ^ *}} \right) = {{\mathcal N}^ * }\left( {0.52,6.37,25.61} \right)$. 

Lachi et al. collected tweet graphs on the vaccination debate in Italy between 2019 and 2022 \cite{Vaccine_Lachi_twitter}. The initial opinions are No-Vaccine supporters $(1)$ and Pro-Vaccine supporters $(0)$. The graph is the retweet graph of Twitter with many multiple links. We clarify the political leanings of agents in the same way as Ref. \cite{Emergence_polarization_PRL} as follows. The opinion of an agent $i$ is computed as ${s_i} = \frac{1}{2}( {{s_0} + \frac{{\sum\nolimits_j {{n_j}{s_j}} }}{n}} )$, where $j$ denotes the neighbors of agent $i$, $n_j$ denotes the number of directed links $\overrightarrow {ij}$ and $n$ denotes the total number of $j$. The opinion $s_i$ is a continuous value from 0 to 1. We denote ${s_i} \in \left( {2/3,1} \right]$ as opinion $+$, ${s_i} \in \left( {1/3,2/3} \right]$ as opinion $\odot$, and ${s_i} \in \left[ {0,1/3} \right]$ as opinion $-$. The fraction of three opinions is $({x_ + }, {x_ \odot }, {x_ - }) = (0.356,0.005,0.639)$ in August 2020. The fraction of nine directed links is $( {{\pi _{\scriptsize\overrightarrow { -  - } }},{\pi _{\scriptsize\overrightarrow { \odot  - } }},{\pi _{\scriptsize\overrightarrow { +  - } }},{\pi _{\scriptsize\overrightarrow { -  \odot } }},{\pi _{\scriptsize\overrightarrow { \odot  \odot } }},{\pi _{\scriptsize\overrightarrow { +  \odot } }},{\pi _{\scriptsize\overrightarrow { -  + } }},{\pi _{\scriptsize\overrightarrow { \odot  + } }},{\pi _{\scriptsize\overrightarrow { +  + } }}} ) = (0.44738,0.0026,0.01226,0.00248,0.00000,0.00119,0.00592,\\0.00384,0.52428)$. Thus, $\left( {{K_0 ^ *},{K_1 ^ *},{K_2 ^ *}} \right) = {{\mathcal N}^ * }\left( {0.55,0.98,25.03} \right)$.

\bibliography{stable_polarization_auto}           

\end{document}